\documentclass{article}
\usepackage[utf8]{inputenc}
\usepackage{spconf,amsmath,graphicx}
\usepackage{cite}
\usepackage{caption}
\usepackage{subcaption}
\usepackage{amsfonts}
\usepackage{amssymb}
\usepackage{amsthm}
\usepackage{gensymb}
\usepackage{xcolor}
\usepackage{url}
\usepackage[ruled,vlined]{algorithm2e}

\usepackage{stfloats}
\usepackage{cases}

\def\ut{U_{mn}(t)}
\def\to{\tau_{mn}^o}
\def\te{\tau_{mn}^e}
\def\df{f_s}
\def\dt{\Delta \tau}
\def\sinc{\text{sinc}}

\title{Time-Modulated Intelligent Reflecting Surface for Waveform Security 
}

\name{$\text{Zhaoyi Xu}$ \thanks{Work supported by NSF under grant ECCS-2033433 and ECCS-2320568 and ARO under grant W911NF2320103 and W911NF2110071.} and $\text{Athina Petropulu}$}
\address{Dept. of Electrical and Computer Engineering, Rutgers University\\}
\begin{document}
%
\setlength{\abovedisplayskip}{5pt}
\setlength{\belowdisplayskip}{5pt}
\maketitle
%

\begin{abstract}
We consider an OFDM transmitter aided by an intelligent reflecting surface (IRS) and propose a novel approach to enhance waveform security by employing time modulation (TM) at the IRS side.
By controlling the periodic TM pattern of the IRS elements, the system is designed to preserve communication information towards an authorized recipient and scramble the information towards all other directions. 
We introduce two modes of TM pattern control: the linear mode, in which we design common TM parameters for entire rows or columns of the IRS, and the planar mode, where we design TM parameters for each individual IRS unit. Due to the required fewer switches, the linear mode is easier to implement as compared to the planar mode. However, the linear model results in a beampattern that has sidelobes, over which the transmitted information is not sufficiently scrambled. We show that the sidelobes of the linear mode can be suppressed by exploiting the high diversity available in that mode.
\end{abstract}

\vspace{-2mm}
\section{Introduction}
\vspace{-1mm}

An Intelligent Reflecting Surface (IRS) is a passive array comprised of printed dipole elements, each capable of dynamically adjusting the phase of impinging electromagnetic waves through computer-controlled means \cite{Wu2019intelligent, Wu2020towards}. These elements can work collaboratively to achieve beamforming in desired directions and nullify signals in undesired directions. One of the key advantages of the IRS is the absence of a radio-frequency (RF) chain, resulting in minimal hardware costs and power consumption. This is coupled with the substantial beamforming gain derived from the numerous reflecting units present within the surface. Additionally, the IRS can reconfigure the wireless propagation environment by providing extra virtual line-of-sight (LoS) paths.
Due to those advantages, IRS has been studied for enhancing communication performance~\cite{hua2021IRS_comm,xie2021IRS_comm}, and aiding radar sensing~\cite{Buzzi2022IRS_radar,shao2022IRS_radar,chen2021IRS_radar}.
The IRS has also been explored for enhancing physical layer security (PLS) and preventing eavesdropping by malicious users~\cite{cui2019IRS_secure,chen2019IRS_secure,yang2021IRS_secure,Hua2023IRS_security}. 

Directional modulation (DM) is a promising PLS approach that preserves the communication information in a certain direction while distorting it in all other directions. Eavesdroppers attempting to intercept data streams, when positioned outside the designated direction, encounter scrambled data.
Notably different from  PLS approaches such as cooperative relaying strategies~\cite{Dong2010Improving,Li2020relay,Li2011cooperative} and artificial noise~\cite{zhang2019AN,wang2017AN}, DM operates without the need for channel state information (CSI), and without generating interference to the legitimate receiver. 
One way to achieve DM is via feeding a time-modulated array (TMA) with an orthogonal frequency-division multiplexing (OFDM) signal~\cite{Ding2019time_modulated}. In a TMA, the antennas connect and disconnect to the RF chain in a periodic manner which gives rise to harmonic signals with controllable amplitude, phase, and frequency~\cite{kummer1963TMA}. By taking the on/off pattern period to be equal to the OFDM symbol duration, the harmonic frequency is equal to the OFDM subcarrier spacing, and the resulting harmonic signals cause scrambling of the data symbols. The scrambled symbol on each subcarrier is a weighted combination of the original data symbols of all subcarriers. The harmonics towards a certain direction can be eliminated by the appropriate design of the TM parameters (``on" time instants and ``on" time duration) and the applied antenna weights.


In this paper, we propose a secure system that realizes DM by implementing time modulation (TM) at the IRS instead of the transmitter. We will refer to the proposed system as time-modulated IRS (TM-IRS).
In TM-IRS, the IRS units are activated in a periodic fashion, and the activation parameters are designed to achieve DM in the 3D space where only the receiver at a certain azimuth and elevation direction can receive the original data symbols. 
Two modes of TM parameter control schemes are proposed, namely, the linear mode and the planar mode. In the linear mode, harmonics are mitigated in a specific direction by configuring TM parameters for the columns or rows of the IRS. Each row/column could share identical TM parameters, but distinct rows/columns have different parameters. In the planar mode, distortion management is achieved by setting unique TM parameters for each IRS unit. In all cases, the TM parameters are provided in closed form.
In the linear mode, each IRS row/column can be controlled by one switch, which makes it a simpler structure. 
However, the linear mode gives rise to high sidelobes over which the transmitted information is not sufficiently scrambled.
We show that one can suppress the sidelobe level by exploiting the available diversity that derives from the independent design of each column or row.
Compared with the time-modulated transmitter in \cite{Ding2019time_modulated}, the substantial beamforming gain from the IRS compensates for the power loss due to the deactivation of radiating elements during the use of TM.

\smallskip
\noindent \textit{Related literature-} Using TM-IRS for securing communication information has been explored  in~\cite{Sedeh2022TMIRS_security}, but in a different context than our contribution. In ~\cite{Sedeh2022TMIRS_security}, the TM parameters are changed in a pseudorandom fashion giving rise to a noise-like spectrum with near-zero power spectral density, and thus preventing the eavesdropper from detecting the communication link. By using the same pseudorandom key generator, the legitimate recipient can successfully demodulate the reflected signal. So, \cite{Sedeh2022TMIRS_security} focuses on the covertness of the signal and on increasing the missed detection probability of eavesdropping, and relies on a shared key between the IRS and the legitimate user. Our proposed approach aims to preserve the communication information in a certain direction while distorting it in all other directions. 



\vspace{-3mm}
\section{Time-Modulated Intelligent Reflecting Surface}
\begin{figure}
    \centering
    \includegraphics[width = 5cm]{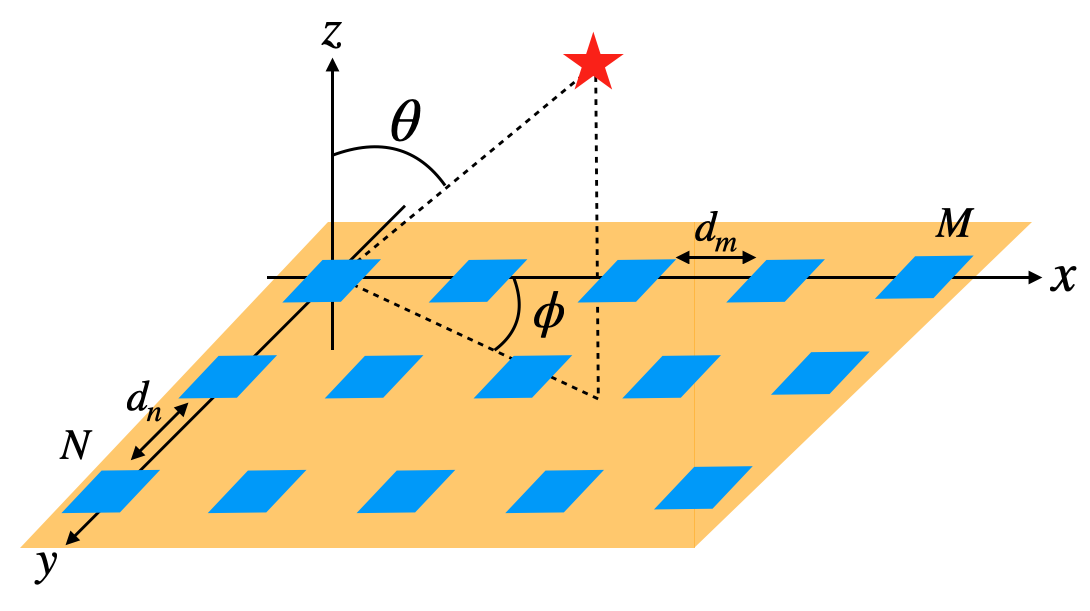}
    \vspace{-5mm}
    \caption{{\small An IRS example.}}
    \vspace{-5mm}
    \label{fig:IRS}
\end{figure}

\vspace{-1mm}
Consider an IRS with $N \times M$ passive units, as shown in Fig.~\ref{fig:IRS}, aiding a uniform linear array (ULA)  transmitter with $K$ elements spaced by $d_t$. 
The signal that arrives at the $mn$-th element of IRS can be expressed as
 {\begin{equation}
    x_{mn}(t) = a_{mn}(\theta_T, \phi_T)\beta \mkern-4mu \sum_{k=0}^{K-1}e^{-j2\pi kd_t \frac{\sin{\theta_I}}{\lambda}}w_k e_k(t),
    \label{eq:x_mn}
\end{equation}
}
where $\lambda$ is the wavelength of carrier frequency; $a_{mn}(\theta,\phi) = e^{-j2\pi/\lambda (md_m\sin{\theta}\cos{\phi}+ nd_n\sin{\theta}\sin{\phi})}$ 
is the far-field array factor of the $mn$-th element~\cite{Yurduseven2020TM_IRS}; $d_m$ and $d_n$ are respectively the unit spacing along the $x$-axis and $y$-axis and both equal to $\lambda/2$; $\theta_T$ and $\phi_T$ are respectively the elevation and azimuth angle of the ULA transmitter w.r.t. the IRS; $\beta$ denotes the complex path loss; $\theta_I$ is the direction of IRS in the view of transmitter; $w_k$ and $e_k(t)$ are respectively antenna weight and baseband signal of the $k$-th transmit antenna. Of course, there is also additive noise, which we skip in the expressions but we still add it in our simulations. We will certainly study noise effects in our future work.
In the above, the IRS from the point of view of the transmitter is a point target. This is because the units of an IRS are very small, and even with a large number of units, the IRS has a small size~\cite{Hua2023IRS_security}.

Each IRS unit is connected to a high-speed switch and a phase shifter. For simplicity, here we consider single-pole-single-throw (SPST) switches, i.e., the switches have two states: ``on" and ``off". Let the on/off activation pattern of the $mn$-th unit be denoted by $\ut$, and the repeating frequency of the activation pattern by $\df$. 
Let $t_{mn}^o$ and $t_{mn}^e$ respectively denote the ``turn-on" and ``turn-off" time of the $mn$-th unit. For simplicity, we use normalized ``on" and ``off" time instants, $\to = t_{mn}^o\df  \in [0,1]$ and $\te = t_{mn}^e\df \in [0,1]$.
Let $\Delta\tau_{mn} = \te - \to$ be the normalized ``on" time duration of the $mn$-th unit. 
We take $\to< \te$,  and set $U_{mn}(t)$ to be $1$ only when $t\df\in[\to, \te]$; otherwise it is $0$.
Based on this activation pattern, the periodic square waveform $U_{mn}(t)$  can be expressed as Fourier series, i.e., 
\begin{multline}
    \ut 
    = \sum_{h=-\infty}^{\infty}e^{j2\pi h\df t} \dt_{mn}\sinc(h\pi\dt_{mn})\\ \times e^{-jh\pi(2\to+\dt_{mn})},
    \label{eq:ut}
\end{multline}
where $\sinc(\cdot)$ denotes the sinc function. It can be seen that the square-shaped activation pattern gives rise to harmonics whose frequency is determined by the repeating frequency of the pattern. 
Let the transmitter operate as a phased array (thus $e_k(t) = e(t), \forall k$), transmitting an OFDM signal with $N_s$ subcarriers spaced at $\df$. Over the duration of $N_p$ OFDM symbols, each antenna is fed with the signal
\begin{equation}
    e(t) = \sum_{\mu=0}^{N_p-1}\sum_{i=0}^{N_s-1} d(i,\mu) e^{j2\pi (f_c+i f_s) t} \text{rect}(\frac{t-\mu T_p}{T_p}),
    \label{eq:baseband}
\end{equation}
where $d(i,\mu)$ is the data symbol on the $i$-th subcarrier during the $\mu$-th OFDM symbol, $f_c$ is the carrier frequency, $T_p$ is the OFDM symbol duration and $\text{rect}(t/T_p)$ denotes a rectangular pulse of duration $T_p$.
On setting $w_k = e^{j2\pi k d_t \frac{\sin\theta_I}{\lambda}}$, the signal transmitted by all antennas adds up coherently at the TM-IRS.

The signal  reflected by the IRS towards  $(\theta,\phi)$ equals
\begin{equation}
\begin{aligned}
    y(t,\theta,\phi) &= \sum_{m=0}^{M-1}\sum_{n=0}^{N-1}a_{mn}(\theta,\phi)c_{mn}\ut x_{mn}(t)
    ,
\end{aligned}
\label{eq: transmitted signal}
\end{equation}
where $c_{mn}$ is the unit modulus weight applied by the $mn$-th IRS unit.

Since the data symbols are independent between OFDM symbols, in the following analysis we will only focus on one OFDM symbol, and discard the rectangular function. 
On substituting \eqref{eq:x_mn}, \eqref{eq:ut}  and \eqref{eq:baseband} into \eqref{eq: transmitted signal}, we get
\begin{align}
     y(t,\theta,\phi) &= \beta K \sum_{i=0}^{N_s-1}d(i)e^{j2\pi (f_c + i f_s) t}\nonumber\\
    &\times \sum_{h=-\infty}^{\infty} e^{j2\pi h f_s t} \sum_{m=0}^{M-1}\sum_{n=0}^{N-1} B(h,\Omega_{mn}, \theta,\phi),
    \label{eq: transmitted signal 2}
\end{align}
where $\Omega_{mn} = \{\Delta\tau_{mn},\tau_{mn}^o,c_{mn}\}$ denotes the set of parameters of the $mn$-th unit, and
\begin{multline*}
    B(h,\Omega_{mn}, \theta,\phi) =  a_{mn}(\theta,\phi)a_{mn}(\theta_T,\phi_T)c_{mn}\\\times \Delta \tau_{mn}\sinc(h\pi \Delta\tau_{mn})e^{-jh\pi(\Delta\tau_{mn}+ 2 \tau^o_{mn})}
\end{multline*}
is the coefficient of the $h$-th harmonic generated by the $mn$-th unit at direction $(\theta,\phi)$. 

In \eqref{eq: transmitted signal 2}, each subcarrier of every TM-IRS unit will generate harmonics that overlap with all other subcarriers. As a result, the $i$-th subcarrier contains the weighted summation of the symbols of all subcarriers from all IRS units, where the weights are the complex coefficient $B(h,\Omega_{mn}, \theta,\phi)$. Thus, the data symbol on every subcarrier is scrambled differently.  After scrambling, the data symbol on the $i$-th subcarrier of \eqref{eq: transmitted signal 2} can be  expressed as
\begin{equation}
    d'(i,\theta,\phi) = 
    \beta K \mkern-6mu\sum_{s=0}^{N_s-1} \sum_{m=0}^{M-1}\sum_{n=0}^{N-1}\mkern-2mu d(s)B(i-s,\Omega_{mn},\theta,\phi).
    \label{eq:scrambled symbol}
\end{equation}

\vspace{-4mm}
\section{3D Directional Modulation with TM-IRS}
\vspace{-1mm}
Let $(\theta_c,\phi_c)$ denote the direction of the legitimate user w.r.t. the TM-IRS. The goal is designing the TM parameters so that we eliminate harmonics (and thus avoid scrambling) in the direction of $(\theta_c,\phi_c)$ while preserving the {signal on the fundamental harmonic frequency}. 
We can first let $c_{mn} = [a_{mn}(\theta_c,\phi_c)a_{mn}(\theta_T,\phi_T)]^{-1} = [a_{mn}(\theta_c,\phi_c)a_{mn}(\theta_T,\phi_T)]^*$ (note that $a_{mn}(\theta,\phi)$ is unit-modulus).
Let us set a universal ``on" time duration for all units to simplify the problem, i.e., $\dt_{mn} = \dt\in (0,1), \forall n,m$. 
In the desired direction,  \eqref{eq:scrambled symbol}  can be simplified as
\vspace{-3mm}
\begin{multline}
     d'(i,\theta_c,\phi_c) =
     \beta K \dt  \sum_{s=0}^{N_s-1} d(s)   \sinc[(i-s)\pi \dt] \\\times
     e^{-j(i-s)\pi\dt}
       \sum_{m=0}^{M-1}\sum_{n=0}^{N-1} e^{-j(i-s)2\pi\to}.
      \label{eq: LU}
\end{multline}
In order to eliminate harmonics and thus avoid scrambling towards $(\theta_c,\phi_c)$, we need to enforce the following condition: 
\begin{equation*}
    \sum_{m=0}^{M-1}\sum_{n=0}^{N-1} e^{-j(i-s)2\pi\to} = 0, \forall i \neq s.
\end{equation*} 
This can be satisfied by the following three  sets:
\vspace{-1mm}
{\small
\begin{align}
    &|\tau_{mn}^o-\tau_{mn'}^o| \in \{\frac{1}{N},\frac{2}{N},\dots,\frac{N-1}{N}\},\forall n\neq n',m, \label{eq:column} \\
    &|\tau_{mn}^o-\tau_{m'n}^o| \in \{\frac{1}{M},\frac{2}{M},\dots,\frac{M-1}{M}\},\forall m\neq m', n \label{eq:row} \\
    &|\tau_{mn}^o-\tau_{m'n'}^o| \in \{\frac{1}{MN},\dots,\frac{MN-1}{MN}\}, \forall mn\neq m'n'.
    \label{eq:plane}
\end{align}}%

\noindent\textbf{Linear Mode - }
In the above solution sets, \eqref{eq:column} and \eqref{eq:row}  respectively eliminate the harmonics in the desired direction by operating along each column and row of the TM-IRS; we refer to this as the linear mode. 
In the linear mode, the TM parameters are individually designed on each column/row, thus the ``on" time duration  needs to be identical for each column/row while it could be different between columns/rows, i.e.,
\begin{equation}
    \hspace{-1mm}
    \begin{cases}
        |\tau_{mn}^o-\tau_{mn'}^o| \in \{\frac{1}{N},\frac{2}{N},\dots,\frac{N-1}{N}\},\forall n\neq n', m,\\
        \dt_{mn} = \dt_m, \forall n; \dt_m \neq \dt_{m'}, \forall m\neq m'.
    \end{cases}
\end{equation}%
\begin{equation}
\hspace{-1.5mm}
    \begin{cases}
        |\tau_{mn}^o-\tau_{m'n}^o| \in \{\frac{1}{M},\frac{2}{M},\dots,\frac{M-1}{M}\},\forall m\neq m', n\\
        \dt_{mn} = \dt_n, \forall m; \dt_n \neq \dt_{n'}, \forall n\neq n'.
    \end{cases}
\end{equation}
The minimal ``on" time instants of each column/row could be different, due to the fact that \eqref{eq:column} and \eqref{eq:row}  affect only the difference between ``on" time instants.

The linear mode offers three forms of diversity: (i) different order of $\to$ in each column/row; (ii) different $\dt_{mn}$ between columns/rows; (iii) different minimal ``on" time instants (or say ``on" time offsets) between columns/rows. 

\noindent\textbf{Planar Mode - }
The third solution set, i.e., \eqref{eq:plane}, eliminates the harmonics with the help of all units on the surface, and thus the corresponding TM-IRS operation is referred to as planar mode. As indicated in \eqref{eq:plane}, the ``on" time instants $\to$ do not necessarily need to be in sorted order. In fact, the planar mode can be reduced to the linear mode, when the differences between $\to$ on every column/row are identical. Thus for the planar mode, the ``on" time instants should be properly designed. Note that, in the planar mode, every unit has the same ``on" time duration. 


{The planar mode offers two forms of diversity: (i) the ``on" time offset and (ii) the order of ``on" time instants.} 

\noindent\textbf{Enhanced Linear Mode - }
Compared with the planar mode, the linear mode can be realized with fewer switches and thus is easier to control and implement.
However,  the linear mode may give rise to more sidelobes than the planar case. Over those sidelobes, the scrambling is not that severe and the data symbols can still be extracted. To tackle this issue, one can leverage the aforementioned diversity and change the assigned TM parameters between several OFDM symbols. Since the coefficients of harmonics are determined by the TM parameters, the scrambling and thus the location of sidelobes will be altered for different sets of parameters.
Even if the eavesdropper is located in the DM sidelobe for certain OFDM symbols, it will receive scrambled data symbols with wrong information in other OFDM symbols. 

If the parameters are changed in a random fashion between OFDM symbols, or blocks of OFDM symbols, the eavesdropper will not be able to learn the used TM pattern. If $M,N$ are large (the IRS is large) and/or there is a large number of OFDM subcarriers, it would be computationally prohibitive for the eavesdropper to find those parameters exhaustively, as the scrambling is generated differently on all subcarriers and all IRS units.

To benefit from the diversity, we need to use different sets of TM parameters periodically.
However, although the TM parameters are given in closed form, as we increase the frequency of updating them, the computation cost increases. This is because we need to choose those parameters so that the same set is not chosen twice.


\vspace{-0mm}
\begin{table}[t]
\centering
\footnotesize{
\caption{\small {System Parameters}}
\vspace{-3mm}
\label{table:system_parameters}

\begin{tabular}{ ||c||c|c|  }
 \hline
 Parameter & Symbol & Value\\
 \hline
 Center frequency   & $f_c$    &$24$ GHz\\
 Subcarrier spacing &   ${\df} $  & $120$ kHz\\
 Number of subcarriers & $N_s$ & 64\\
 Number of OFDM symbols & $N_p$ & $2^{14}$\\
Size of TM-IRS & $M \times N$ & $16 \times 16$\\
 Angle of legitimate user & $(\theta_c,\phi_c)$ & $(40\degree, 30\degree)$ \\
 \hline
\end{tabular}
\label{table:parameters}
}
\end{table}

\vspace{-3mm}
\section{Numerical results}
\vspace{-1mm}
In this section, we demonstrate the bit error rate (BER) performance of the proposed system via simulations. The system parameters are listed in Table~\ref{table:parameters}. During the simulations, the path loss was set to $1$. In a practical case, the phase shift due to path loss can be estimated and calibrated during channel estimation between the transmitter and the IRS. The random binary stream was modulated into data symbols via quadrature phase-shift keying (QPSK).
The ULA transmitter has $8$ half-wavelength antennas and the TM-IRS is at $\theta_I = 30\degree$ w.r.t. the transmitter. The transmitter is located at $(\theta_T, \phi_T) = (15\degree, 10\degree)$ w.r.t. the TM-IRS. The ratio between data symbol power and additive noise power was set to $0$ dB. Unless specified, the normalized ``on" time duration was set to $0.7$.
The simulation results are plotted on a logarithmic scale, and for the purpose of plotting,  a small offset of $1^{-10}$ was added to all results. In the figures, the dark spots represent areas of very low BER.

\begin{figure}
    \centering
    \begin{subfigure}[b]{0.235\textwidth}
        \centering
        \includegraphics[width= 4.2cm]{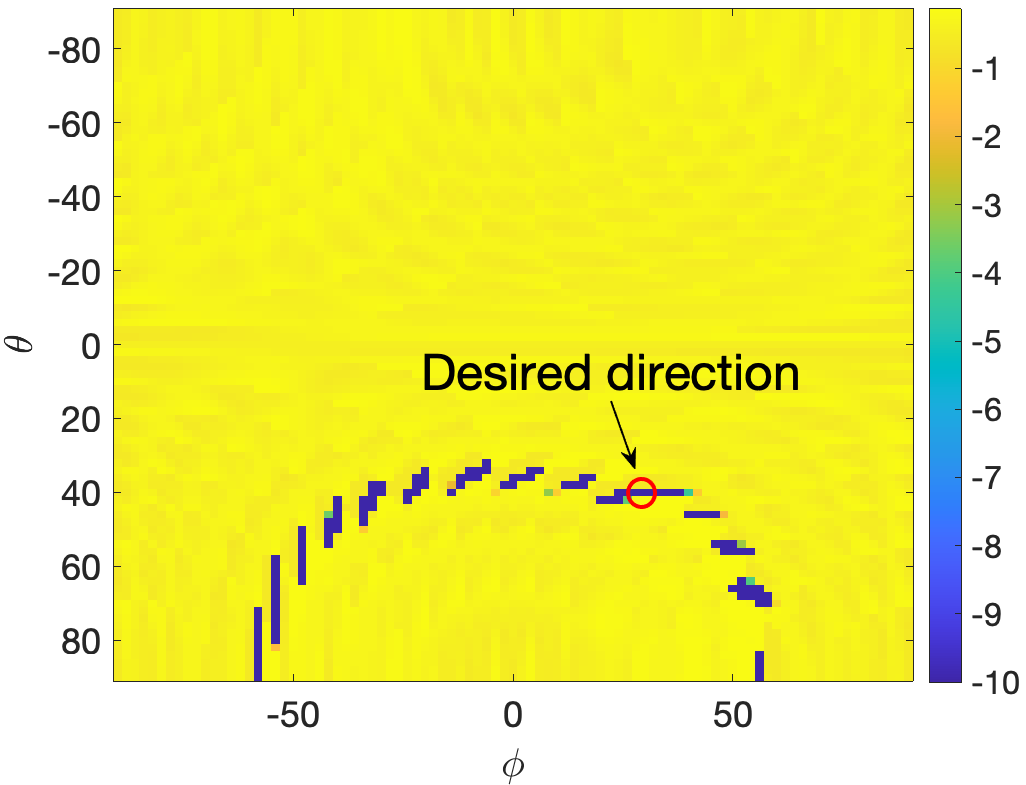}
        \vspace{-6mm}
        \caption{Linear mode (column)}
        \label{fig:column}
    \end{subfigure}
    \hfill
    \begin{subfigure}[b]{0.235\textwidth}
        \centering
        \includegraphics[width= 4.2cm]{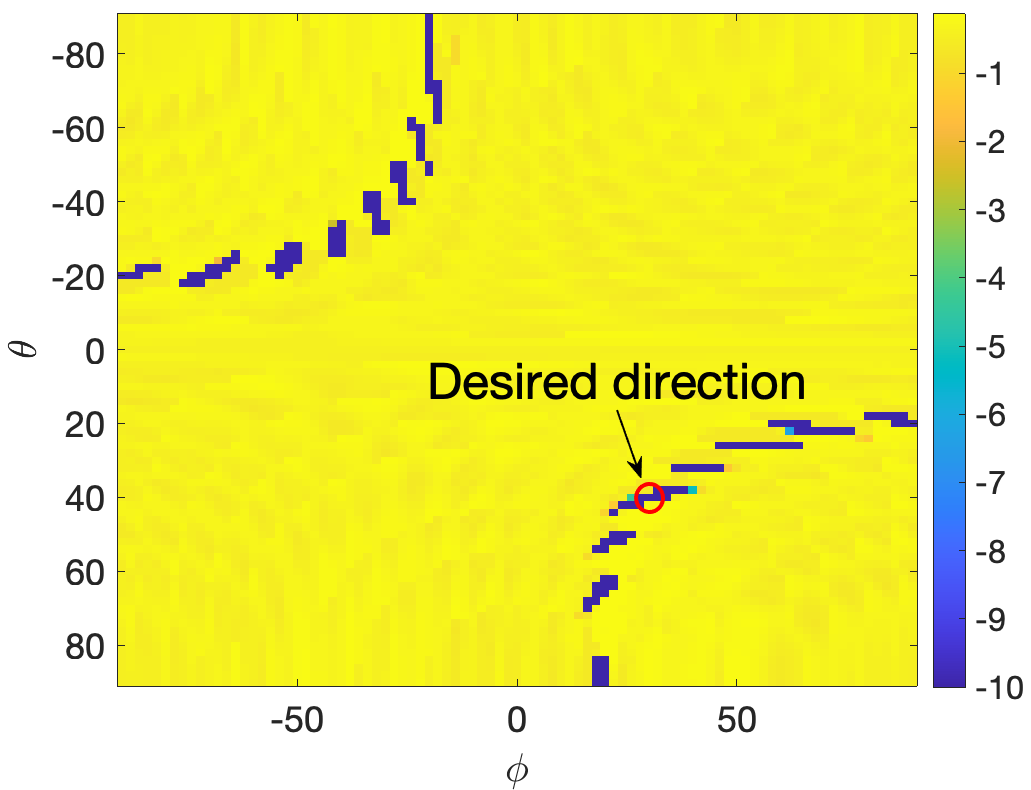}
        \vspace{-6mm}
        \caption{Linear mode (row)}
        \label{fig:row}
    \end{subfigure}
    \label{fig:linear}
    \vspace{-3mm}
    \caption{\small{BER of the linear mode, where the harmonics are canceled by operating along (a) each column and (b) each row of the IRS.} 
    }
    \vspace{-3mm}
\end{figure}

\smallskip
\noindent{\it Linear mode -}
Here, the ``on" time instants along each column/row of the TM-IRS were taken in random order and were kept the same between columns/rows. 
The same ``on" time duration was used in all cells.
The TM parameters remained unchanged between OFDM symbols. 
The BER results of canceling harmonics by operating along the columns and rows of the TM-IRS are respectively shown in Figs.~\ref{fig:column} and  ~\ref{fig:row}.
As one can see, the BER is very low in the desired direction of $(\theta_c,\phi_c) = (40\degree, 30\degree)$. However, there are also areas with low BER away from the desired direction, due to insufficient scrambling. If an eavesdropper was to be positioned there, it could successfully decipher the data symbols.
Also, from those two figures, one  can see that 
operating along different IRS dimensions results in different sidelobe behavior.

\begin{figure}
    \centering
    \begin{subfigure}[b]{0.235\textwidth}
        \centering
        \includegraphics[width= 4.2cm]{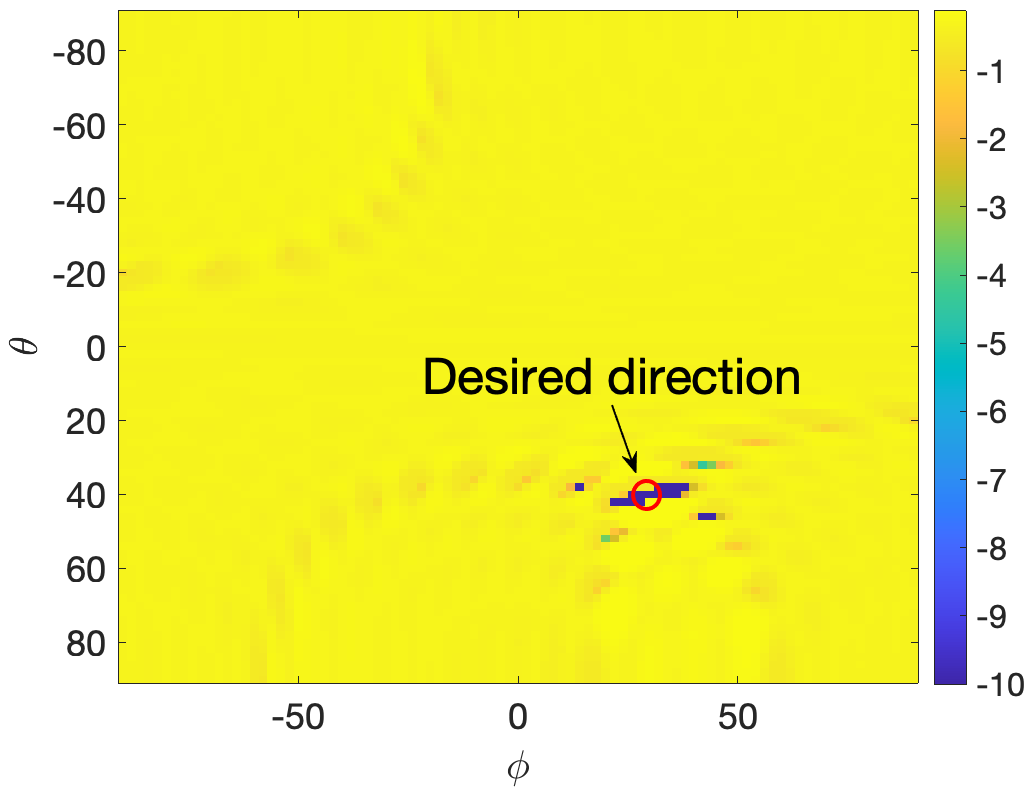}
        \vspace{-6mm}
        \caption{\small{Planar mode}}
        \label{fig:planar}
    \end{subfigure}
    \hfill
    \begin{subfigure}[b]{0.235\textwidth}
        \centering
        \includegraphics[width= 4.2cm]{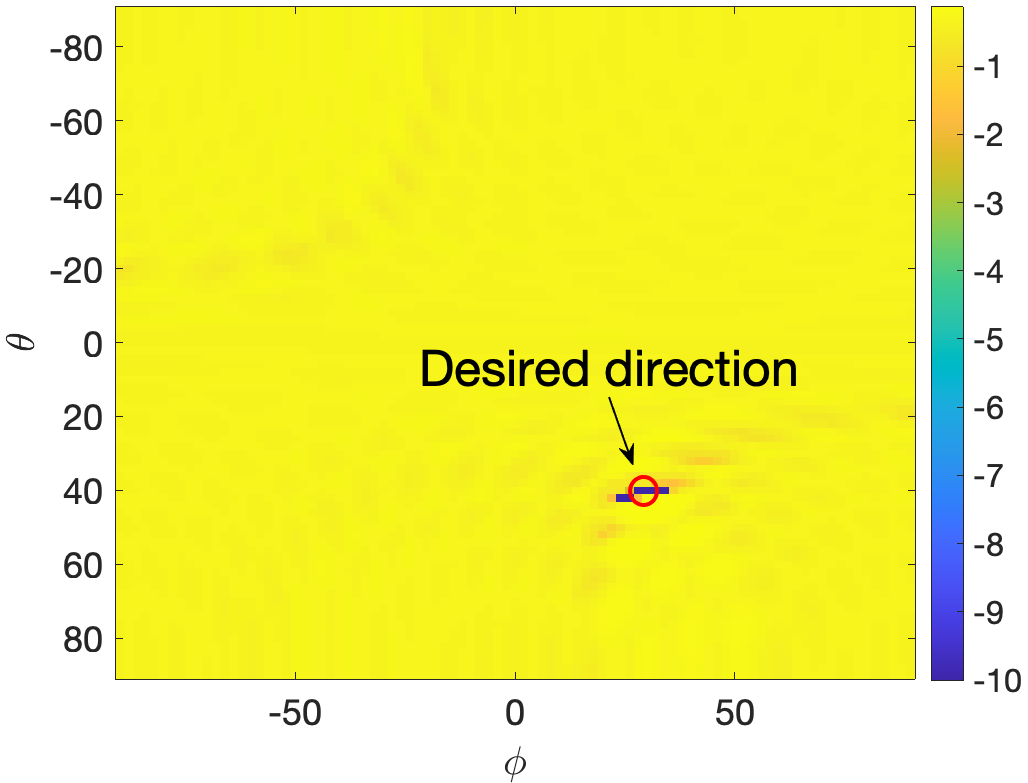}
        \vspace{-6mm}
        \caption{\small{Enhanced linear mode }}
        \label{fig:col_new}
    \end{subfigure}
    \vspace{-3mm}
    \caption{{\small BER of (a) the planar mode, where the harmonics are eliminated by operating on the entire IRS surface; (b) the enhanced linear mode where the TM parameters change every $256$ OFDM symbols.}}
    \vspace{-5mm}
\end{figure}

\smallskip
\noindent{\it Planar mode - }
Here, the ``on"  time instants are randomly assigned to the units and remain unchanged between OFDM symbols. Note that the random order of the ``on"  time instants
improves the performance of the planar mode.
The achieved BER  is shown in Fig.~\ref{fig:planar}, where one can see that the sidelobes are significantly reduced. It is interesting to notice that there exist two faint traces of relatively low BER, which are very similar to the patterns of the linear mode. The area with $0$ BER is exactly the intersection area of the two faint traces.

\smallskip
\noindent{\it Enhanced Linear mode -}
Here, the TM-IRS operates in linear mode (column), and the ``on" time duration of each column was randomly generated and changed every $256$ OFDM symbols. All other parameters are the same as  in the first experiment.
The achieved BER is shown in Fig.~\ref{fig:col_new}, where one can see that the sidelobes are fully suppressed and the area with low BER is smaller than that in Fig.~\ref{fig:column}. Each time the ``on"  time duration was changed, the coefficients of harmonics were altered and so did the sidelobes. 
As a consequence, the BER here is effectively the average of $2^{14}/256 = 64$ different BER patterns. 

It is also possible to suppress the sidelobes by using different ``on" time instant order between columns/rows, and also by changing them between OFDM symbols. Due to the limited space, we cannot show results here.

\vspace{-3mm}
\section{Conclusion}
\vspace{-1mm}
We have proposed a novel secure OFDM transmitter aided by a TM-IRS, that preserves the information in the direction of the legitimate receiver and makes it appear scrambled in all other directions in the 3D space. To realize such 3D DM, the IRS elements are tuned on/off in a periodic manner, with a period equal to the OFDM symbol duration.  
We have studied two different working modes,  where the harmonics are eliminated by operating either along each column/row of the IRS (linear mode) or on the entire IRS surface (planar mode). The linear mode exhibits more sidelobes than the planar mode, representing areas of low BER away from the desired direction, but offers more diversity than the planar mode.
We have shown how the sidelobes can be reduced by exploiting that diversity, namely, by assigning different ``on" time durations to columns/rows and changing them every several OFDM symbols. 

\bibliographystyle{IEEE}
\bibliography{ref.bib}

\end{document}